\documentclass[graybox]{svmult}

\usepackage{mathptmx}       
\usepackage{helvet}         
\usepackage{courier}        
\usepackage{type1cm}        
\usepackage{makeidx}         
\usepackage{graphicx}        
\usepackage{multicol}        
\usepackage[bottom]{footmisc}

\usepackage{natbib}

\newcommand{\kms}{{\rm km~s^{-1}}}
\newcommand{\kpc}{{\rm kpc}}
\newcommand{\kmskpc}{\kms\kpc^{-1}}

\newcommand{\Ms}{M_\odot}
\newcommand{\degree}{\circ}

\makeindex             

\begin{document}
\title*{Theoretical Models of the Galactic Bulge}
\author{Juntai Shen \& Zhao-Yu Li}
\authorrunning{Shen \& Li}
\institute{Key Laboratory for Research in Galaxies and Cosmology, Shanghai Astronomical Observatory,
Chinese Academy of Sciences, 80 Nandan Road, Shanghai 200030, China. \email{jshen,lizy@shao.ac.cn}}
%
%
\maketitle

\abstract{Near infrared images from the \textit{COBE} satellite presented the first clear evidence that our Milky Way galaxy contains a boxy shaped bulge.  Recent years have witnessed a gradual paradigm shift in the formation and evolution of the Galactic bulge. Bulges were commonly believed to form in the dynamical violence of galaxy mergers. However, it has become increasingly clear that the main body of the Milky Way bulge is not a classical bulge made by previous major mergers, instead it appears to be a bar seen somewhat end-on. The Milky Way bar can form naturally from a precursor disk and thicken vertically by the internal firehose/buckling instability, giving rise to the boxy appearance. This picture is supported by many lines of evidence, including the asymmetric parallelogram shape, the strong cylindrical rotation (i.e., nearly constant rotation regardless of the height above the disk plane), the existence of an intriguing X-shaped structure in the bulge, and perhaps the metallicity gradients. We review the major theoretical models and techniques to understand the Milky Way bulge. Despite the progresses in recent theoretical attempts, a complete bulge formation model that explains the full kinematics and metallicity distribution is still not fully understood. Upcoming large surveys are expected to shed new light on the formation history of the Galactic bulge.}

\section{A brief overview on the properties of the Galactic bulge}
\label{sec:overview}
Most spiral galaxies consist of three main components, an invisible dark matter halo, an embedded flat disk, and a central bulge. The Milky Way is no exception. The Milky Way bulge comprises about 15\% of the total luminosity, and its stellar mass is about $1.2-1.6\times10^{10}\Ms$ \citep{portai_etal_15a}. Galactic bulges contain crucial information about the galaxy formation and evolution. Major mergers between galaxies generally create a significant classical bulge that is similar to ellipticals in many aspects (see Chapter 6.1 of this book), whereas the long term internal secular evolution in the disk galaxy tends to build up a pseudobulge \citep{kor_ken_04}. Understanding the structure of our Milky Way bulge is nontrivial, mostly because of our location in the disk plane and the severe dust extinction in the optical band.  One the other hand, a huge advantage being inside the Milky Way is the power to resolve and observe individual stars. Here we briefly summarize the structure, chemical composition, age, and kinematics of the Galactic bulge. More detailed reviews on the observational properties of the Galactic bulge can be found in \citet{zoccal_10,rich_13,origli_14} and Chapter 4.1 of this book.

\textbf{Structure.} The presence of a triaxial structure in the inner Galaxy was first hinted from gas kinematics \citep{devauc_64, binney_etal_91, bur_lis_93}. The near infrared images from the \textit{COBE} satellite revealed clearly that the Milky Way contains an asymmetric parallelogram-shaped boxy bulge in the center \citep{weilan_etal_94}. The asymmetry may be explained by a tilted bar; the near end of the bar is closer to us than the far side, consequently it appears to be bigger than the other side \citep{bli_spe_91}. Measurements of the three-dimensional density distribution of the Galactic bulge using various stellar tracers give a triaxial bar with the semi-major axis $\sim 3-4\;\kpc$ and tilted by $\sim 20^\degree-30^\degree$ between the Sun-Galactic Center (GC) line \citep[i.e.,][]{stanek_etal_97,bis_ger_02,ratten_etal_07b,robin_etal_12,cao_etal_13,weg_ger_13,pietru_etal_14}. This main triaxial bar is sometimes dubbed as ``the (boxy) bulge bar'' (i.e., boxy pseudobulge in the notation of this book), whose properties still differ among different research groups.

The existence of other bar structures than the main bulge bar is still under active debate. Based on stars counts from the GLIMPSE (Galactic Legacy Mid-Plane Survey Extraordinaire) survey, \cite{benjam_etal_05} argued for another planar long bar passing through the GC with half-length 4.4 kpc tilted by $\sim 45^\degree$ to the Sun-GC line (see also \citealt{cabrer_etal_07}).
If this long bar is confirmed, then its co-existence with the similarly-sized bulge bar is dynamically puzzling, as their mutual torque tends to align the two bars on a short timescale. Also both observations of external galaxies \citep{erw_spa_02, erw_spa_03} and simulations of long-lived double barred galaxies \citep{deb_she_07,she_deb_09} have shown that the size ratio of two bars is generally about $0.1-0.2$. Using numerical simulations, \cite{mar_ger_11} showed that the observational signatures of the planar long bar may actually be reproduced with a single bar structure. They suggested that the long bar may correspond to the leading ends of the bulge bar in interaction with the adjacent spiral arm heads. \citet{nishiy_etal_05, gonzal_etal_11b} also suggested the existence of a possible secondary inner bar, as the slope of the longitudinal magnitude peak profile of red clumps flattens at $|l|\sim4^\degree$, deviating from the main bar. With the same single barred $N$-body model, \cite{ger_mar_12} showed that such a slope change may be caused by a transition from highly elongated to nearly axisymmetric isodensity contours in the inner boxy bulge.

\begin{figure}
\centerline{\includegraphics[width=.8\hsize]{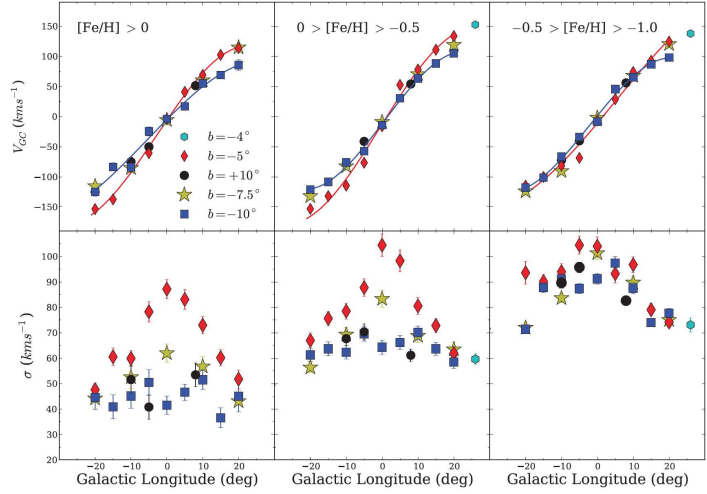}}
\caption{Rotation and velocity dispersion profiles in the ARGOS observations towards the Galactic bulge fields from \cite{ness_etal_13a}. The three columns correspond to three different metallicity bins, decreasing from left to right. Different symbols represent stars in different fields. \textit{Reproduced from \citet{ness_etal_13a}.}}
\label{fig:ness_kine}
\end{figure}

\textbf{Chemical composition and stellar age.} The chemical composition and the stellar age are crucial parameters to constrain galaxy formation history and to establish the connection to stellar populations in other structural components. The bulk of bulge stars is old with a wide range of metal abundances \citep{mcw_ric_94, zoccal_etal_08}, including some of the oldest stars in the Milky Way \citep[e.g.][]{howes_etal_14,sch_cas_14}. The metallicity distribution in the bulge displays a vertical gradient both on and off the minor axis away from the Galactic plane \citep{zoccal_etal_08,gonzal_etal_11a,johnso_etal_11, johnso_etal_12,johnso_etal_13}, while \cite{rich_etal_12} found no major vertical abundance gradient close to the disk plane ($b\le 4^\degree$). \cite{ness_etal_13a} found that stars with $\rm [Fe/H] > -0.5$ are part of the boxy bar/bulge, and the metal-poor stars are likely associated with thick disk.  As shown in Fig.~\ref{fig:ness_kine}, although the rotation curves are similar for three different metallicity bins, the metal-poor population ($-0.5 < {\rm [Fe/H]} < -1.0$) clearly has higher velocity dispersion and spheroidal kinematics. In \citet{ness_etal_13a}, stars with $\rm [Fe/H] \sim +0.15$ are more prominent close to the plane than the metal-poor stars, appearing as a vertical abundance gradient of the bulge.
Most bulge metal-poor stars show enhanced alpha elements compared to both thin and thick disk stars \citep{johnso_etal_11, gonzal_etal_11a, rich_etal_12}, suggesting a rapid formation timescale with respect of both disk components.

Compared to metallicity, age determination is more difficult and imprecise (see the review by \citealt{soderb_10} and references therein). The bulk of bulge stars is old ($\sim 10$ Gyr) \citep[e.g.,][]{ortola_etal_95,lecure_etal_07,clarks_etal_08}. However, intermediate-age metal-rich stars are also detected in the bulge region, with the exact relative fraction still under debate \citep{clarks_etal_11, bensby_etal_11, bensby_etal_12,ness_etal_14}. In addition, there is a nuclear disk of much younger stellar population with ongoing star formation in the central 200 pc (sometimes termed ``nuclear bulge"), whose mass is about $1.5\times 10^9 \Ms$ \citep{launha_etal_02}.

\textbf{Kinematics.} To study systematically the stellar kinematics, \cite{rich_etal_07} initiated the Bulge Radial Velocity Assay (BRAVA) project with M giants as tracers covering the whole Galactic bulge ($\sim$9000 stars). The BRAVA survey revealed clear cylindrical rotation of the bulge, i.e., nearly constant rotation regardless of the height above the disk plane \citep{howard_etal_08, howard_etal_09, rich_etal_08, kunder_etal_12}. BRAVA kinematics also put the Galactic bulge close to the oblate isotropic rotator line in $V_{\rm max}/\sigma-\epsilon$ diagram \citep{binney_78}, distinct clearly from a hot slowly-rotating system like the Milky Way halo supported by velocity dispersion. The BRAVA radial velocity distribution was well reproduced by a self-consistent $N$-body model of a pure-disk Galaxy by Shen et al. (2010), with no need for a significant classical bulge (see Section~\ref{sec:bar}).

The Abundance and Radial velocity Galactic Origins Survey (ARGOS) obtained radial velocities and stellar parameters for 28000 stars in the bulge and inner disk of the Milky Way galaxy across latitudes of $b$ = $-5^\degree$ and $-10^\degree$ \citep{freema_etal_13}. The cylindrical rotation of the bulge was also confirmed in the ARGOS data \citep{ness_etal_13a}. They found a kinematically distinct metal-poor population ($\rm [Fe/H] < -1.0$), which may be related to the thick disk or halo. In the commissioning observation of APO Galactic Evolution Experiment (APOGEE) towards the Galactic bulge region, a high Galactocentric velocity ($V_{\rm GSR} \sim +200$ km s$^{-1}$) and cold ($\sigma \sim 30$ km s$^{-1}$) stream was reported and suggested as the bar supporting orbits \citep{nideve_etal_12}. However, \citet{li_etal_14} suggested that typical bar models do not generate a distinct cold high velocity stream observed from the solar perspective. A smooth high velocity shoulder instead does exist in many bulge fields; it roughly corresponds to the tangential point between the line-of-sight and the bar-supported orbits as shown in the distance-velocity diagram. The cold high velocity stream, if confirmed, may not be necessarily associated with the bar, but could be due to other substructures. Recent Giraffe Inner Bulge Survey (GIBS) observed 24 Galactic bulge fields and confirmed the cylindrical rotation and the lack of a significant cold high velocity peak in the radial velocity distribution \citep{zoccal_etal_14}. Besides the radial velocity, the proper motion is also an important parameter \citep[e.g.,][]{ratten_etal_07a}. Given the stellar distance, the transverse velocities can be estimated. In the Baade's window, the velocity ellipsoid of metal-rich stars shows a vertex deviation in the radial versus transverse velocity, consistent with the bar supporting orbits \citep{soto_etal_07}.

Based on the radial velocity and the [Fe/H] measurement of three minor axis fields, \cite{babusi_etal_10} identified two distinct population: the metal-rich population with bar-like kinematics and the metal-poor population corresponding to an old spheroid or a thick disk \citep[also see][]{hill_etal_11}. The metal-rich population demonstrates smaller velocity dispersion and lower alpha-element enhancement compared to the metal-poor population \citep{johnso_etal_11, uttent_etal_12}. Across the bulge fields in ARGOS survey, the metal-poor population ($\rm [Fe/H] < -1.0$) was also found to be kinematically distinct with large velocity dispersion and non-cylindrical rotation \citep{ness_etal_13b}.

\textbf{X-shaped structure.} A recent development in the bulge structural study is the discovery of an intriguing X-shaped structure (Section~\ref{sec:xshape}). Recently, two groups independently reported the bimodal brightness distribution of the red clump (RC) stars, which can be considered a good standard candle \citep{sta_gar_98}, in the Galactic bulge (\citealt{mcw_zoc_10} hereafter MZ10; \citealt{nataf_etal_10}). MZ10 suggested that the bimodality is hard to explain with a naive tilted bar since the line of sight crossing the bar can only result in stars with one distance. One possibility speculated by \cite{nataf_etal_10} is that one RC population belongs to the bar and the other to the spheroidal component of the bulge.  Another puzzling fact is that distances of the bright and faint RCs are roughly constant at different latitudes, which was hard to understand with a naive straight bar. They proposed that these observed evidences can be well explained with a vertical X-shaped structure in
the bulge region. The existence of this particular structure was later verified
by \cite{saito_etal_11}. They found that the X-shaped structure
exists within (at least) $|l| \leq 2^\circ$, and has front-back symmetry.
Around $b = \pm 5^\degree$, two RCs start to merge, due to severe dust extinction and foreground contamination (MZ10; \citealt{weg_ger_13}).
From numerical simulations, as demonstrated in \cite{li_she_12} and \cite{ness_etal_12}, the buckled bar naturally reproduces the observed X-shape properties in many aspects. The X-shape extends to about half the bar length with similar tilting angle as the bar. The observed north-south symmetry of the X-shape indicates that it must have formed at least a few billion years ago \citep{li_she_12}.

\section{A fully evolutionary bar model as the basis of understanding the Galactic bulge}
\label{sec:bar}
Theoretical modeling of the Milky Way bulge made intense use of $N$-body simulations. The basis of a successful Galactic bulge model is a fully evolutionary bar model that developed naturally from the bar instability of a cold massive disk. Here we describe a successful high-resolution bar model developed in \citet[hereafter S10]{shen_etal_10}, which was initially motivated to match the BRAVA stellar kinematic data.
$N$-body bar models to explain the Galactic bulge were already attempted in early studies such as \cite{fux_97,sevens_etal_99}, but little stellar kinematic data were available to constrain their models.

S10 simulated the self-consistent formation of a bar that
buckles naturally into a thickened state, then scaled that model to fit the BRAVA
kinematic data on bulge rotation and random velocities.
BRAVA measured the stellar radial velocities of M-type giant stars
whose population membership in the bulge is well established.  These giants
provide most of the 2 $\mu$m radiation whose box-shaped light distribution
motivates bar models.  S10 used nearly 5,000 stellar radial velocities in two strips at
latitude $b=-4^\circ$ and $b=-8^\circ$ and at longitude $-10^\circ$ $< l <$
$+10^\circ$, and a strip along the minor axis ($l \equiv 0^\circ$). The strong cylindrical rotation was found in the preliminary BRAVA data \citep{howard_etal_09} consistent with an edge-on, bar-like pseudobulge \citep{kormen_93, kor_ken_04}, although a precise fit of a bar model to the
data was not available. The success prompted S10 to construct a fully
evolutionary $N$-body model that can fit the radial velocity data of BRAVA.

\begin{figure}
\centerline{\includegraphics[angle=-90.,width=0.55\hsize]{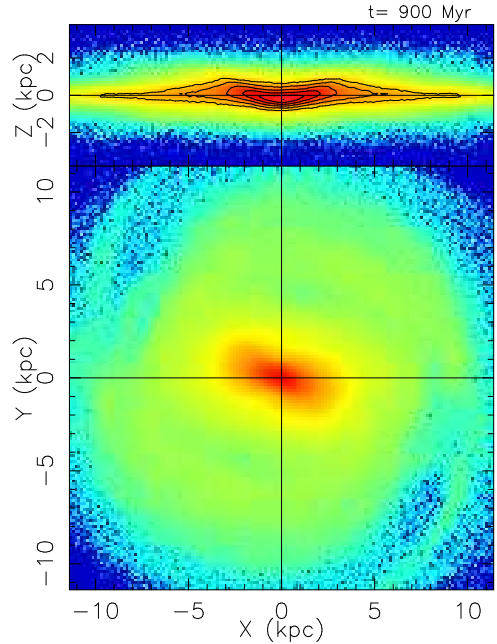}}
\caption{Shortly after its formation, the bar becomes vulnerable to the vertical firehose/buckling instability \citep{toomre_66,raha_etal_91}. The figure illustrates the vigorous buckling instability that makes the initially thin bar bend out of the disk plane, reaching a considerable maximum distortion. After the buckling instability saturates on a short dynamical timescale (in a few hundred million years), the bar is greatly thickened in the vertical direction, giving rise to the boxy shape. (see Fig.~\ref{fig:cont}).}
\label{fig:bending}
\end{figure}

\begin{figure*}[!ht]
\centerline{
\includegraphics[angle=-90.,width=0.45\hsize]{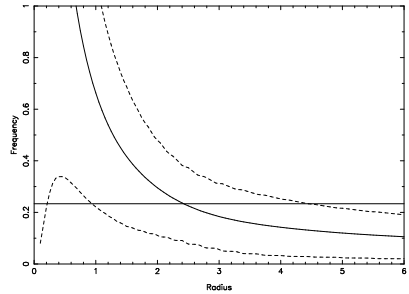}}
\caption{The horizontal line marks the pattern speed $\Omega_p$ of the
  quasi-steady bar in internal simulation unit with $R_{\rm d}=G=M_{\rm
    d}=1$. Here $\Omega_p \approx 39$ km/s/kpc in physical units. The solid
  line shows the curve of the circular angular frequency $\Omega$, and the
  dashed lines mark $\Omega\pm\kappa/2$ at around $t=4.8$ Gyr. \textit{Reproduced from \citet{shen_14} with permission.}}
\label{fig:freq}
\end{figure*}

S10 used a cylindrical particle-mesh code \citep{she_sel_04} to
build fully self-consistent $N$-body galaxies.  The code is highly optimized to study the
evolution of disk galaxies, and it allows S10 to model the disk with at least 1 million
particles to provide high particle resolution near the center where the
density is high.  They tried to construct the simplest self-consistent $N$-body
models that fit the BRAVA data, avoiding contrived models with too many
free parameters.  Initially, they contained only an unbarred disk and a dark
halo.  The profile of the Galactic halo is poorly constrained observationally;
S10 adopted a rigid pseudo-isothermal halo potential which gives a nearly flat initial rotation curve between $5\;\kpc$ and $20\;\kpc$.  A simple halo form allows S10 to run many simulations quickly, which greatly facilitates a parameter search.  A rigid halo also omits dynamical friction on the bar, but the central density of this cored halo is low enough so that friction will be very mild. Since the bulge is embedded well interior to the core radius of the halo, the exact profile of the dark halo at large radii is not critical.

The process of bar formation and thickening in S10 is physically understood. First a bar develops self-consistently via the bar instability from the initially unbarred, thin disk.  Bar formation enhances the radial streaming motions of disk particles, so the radial velocity dispersion quickly grows much bigger than the vertical one.  Consequently the disk buckles vertically out of the plane like a firehose (Fig.~\ref{fig:bending})\footnote{It is puzzling why no galaxy has been caught in the process of violent buckling as shown in Fig.~\ref{fig:bending}. The saturation timescale of the buckling instability is very rapid (about a few hundred Myr), but it is not short enough to miss out the violent buckling phase if one can observe thousands of edge-on barred galaxies. Perhaps this implies that the buckling instability must have happened in the very early assembly stage of disk galaxies, which is much harder to observe.}; this is the well known firehose or buckling instability \citep[e.g.,][]{toomre_66, combes_etal_90,raha_etal_91}.  It raises the vertical velocity dispersion and increases the bar's thickness.  This happens on a short
dynamical timescale and saturates in a few hundred million years.  The central
part of the buckled bar is elevated well above the disk mid-plane and
resembles the peanut morphology of many bulges including the one in our Galaxy.

\begin{figure}
\centerline{
\includegraphics[angle=0.,width=\hsize]{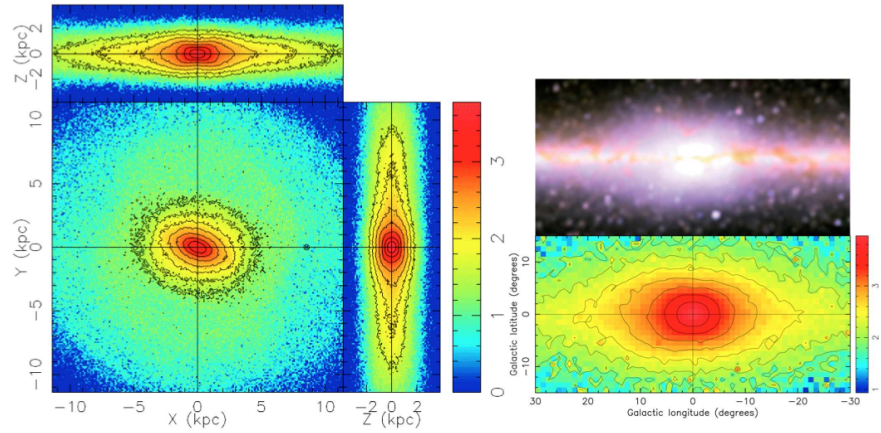}
}
\caption{Left three panels: face-on and side-on views of the surface density
  of our best-fitting model as seen from far away.  The Sun's position 8.5~kpc
  from the Galactic center is marked along the $+x$ axis.  The Galaxy rotates
  clockwise as seen in the face-on projection. Right panels: The COBE DIRBE composite image of the Milky Way
  bulge and model's surface brightness map in solar perspective.  Both the model and observed image show the box-shaped, edge-on bar that appear bigger on its near
  side (positive longitude side).  \textit{The left panel is reproduced from \citet{shen_etal_10} with permission.}}
\label{fig:cont}
\end{figure}

S10 found the one that best matches our BRAVA kinematic data after suitable mass scaling, out of a large set of $N$-body models. The barred disk evolved from a thin exponential disk that contains $M_{\rm d}=4.25\times 10^{10}\Ms$, about 55\% of the total mass at the truncation radius (5 scale-lengths).  The scale-length and scale-height of the initial disk are
$\sim$ 1.9~kpc and 0.2~kpc, respectively. The disk is rotationally supported
and has a Toomre-Q of 1.2.  The amplitude of the final bar is intermediate
between the weakest and strongest bars observed in galaxies. The bar's
minor-to-major axial ratio is about 0.5 to 0.6, and its half-length is $\sim$
4 kpc.  Fig.~\ref{fig:freq} shows the pattern speed of the bar ($\Omega_p \approx 39$
km/s/kpc) and locations of the Lindblad resonances. The bar properties in the best-fitting model are comparable to those obtained in other independent studies. The left panel of Fig.~\ref{fig:cont} shows face-on and side-on views of the projected density of the best-fitting model in S10. A distinctly peanut shaped bulge
is apparent in the edge-on projection.  Fig.~\ref{fig:cont} (right panel) shows the
surface brightness distribution in Galactic coordinates in solar perspective.  Nearby disk stars dilute the peanut shape, but the bar still looks boxy.  The asymmetry in the longitudinal
direction can be understood as a perspective effect; the near end of the bar (at positive Galactic longitude) is closer to the Sun, so it appears bigger and taller than the far side.  Both the boxy shape and the asymmetry are in good agreement with the morphology revealed by the near-infrared image from the COBE satellite
\citep{weilan_etal_94}.

\begin{figure}[!th]
\centerline{
\includegraphics[angle=0.,width=0.85\hsize]{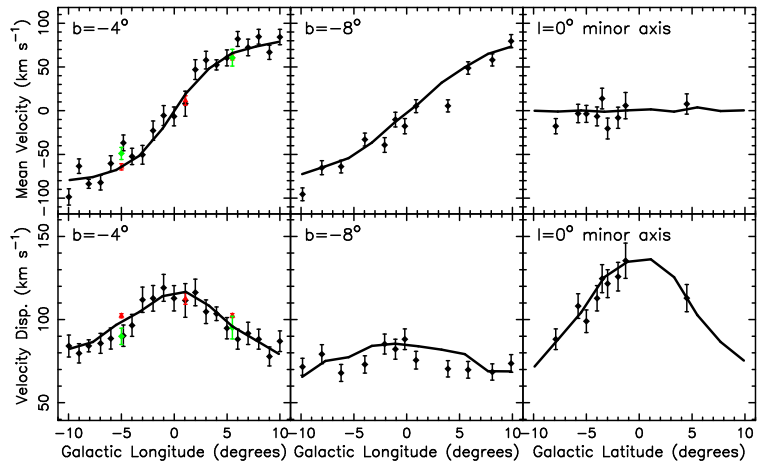}
}
\caption{(top): Mean velocity and velocity dispersion profiles of the
best-fitting model (black lines) compared to all available kinematic
observations.  The left two panels are for the Galactic latitude
$b=-4^\circ$ strip; the middle two panels are for the $b=-8^\circ$
strip; and the right two panels are for the $l=0^\circ$ minor
axis. The black diamonds and their error bars are the BRAVA data; the
green diamonds are for M-type giant stars \citep{rangwa_etal_09}, and the
red triangles are the data on red clump giant stars \citep{rangwa_etal_09}.
This is the first time that a single dynamical model has been compared
with data of such quality.  The agreement is striking. \textit{Reproduced from \citet{shen_etal_10} with permission.}}
\label{fig:velocity}
\end{figure}

Fig.~\ref{fig:velocity} compares the best-fitting model kinematics in S10 (solid lines) with the
mean velocity and velocity dispersion data from the BRAVA and other
surveys.  All velocities have been converted to Galactocentric
values (the line-of-sight velocity that would be observed by a stationary
observer at the Sun's position).  For the first time, this model is able to simultaneously match the mean
velocities and velocity dispersions along two Galactic latitudes ($-4^\circ$
and $-8^\circ$) and along the minor axis. The model comparison with the complete BRAVA data release was also impressive \citep{kunder_etal_12}.

S10 also provided some constraints on the bar angle between the bar and the Sun-GC line. Using both the fit to the velocity profiles and the photometric asymmetry, they found that the overall best-fitting model has a bar angle of $\sim20^\degree$, which agrees reasonably well with other independent studies \citep[e.g.,][]{stanek_etal_97,ratten_etal_07b,cao_etal_13,weg_ger_13}.

The best-fitting model in S10 contains no classical bulge component. S10 also tested whether or not a significant classical bulge is present, since it could have been spun up by the later formation of a bar, flattened thereby and made hard to detect. They found that including a classical bulge with
$>\sim$ 15\% of the disk mass considerably worsens the fit of the model to the data, even if the disk properties are accordingly re-adjusted. If the pre-existing classical bulge is overly massive, then it becomes increasingly hard to match both the mean velocity and velocity dispersions simultaneously \citep[see also][]{saha_etal_12}.

The BRAVA kinematic observations show no sign that the Galaxy contains a significant merger-made,
¡°classical¡± bulge. S10 demonstrated that the boxy pseudobulge is not a separate
component of the Galaxy but rather is an edge-on bar. This result also has important implications for galaxy formation. From a galaxy formation point of view, we live in a pure-disk galaxy. Our Galaxy is not
unusual. In fact, giant, pure-disk galaxies are common in environments like our
own that are far from rich clusters of galaxies \citep{kormen_etal_10,laurik_etal_14}. Classical bulgeless, pure-disk galaxies still present an acute challenge to the current picture of galaxy formation in a universe
dominated by cold dark matter; growing a giant galaxy via
hierarchical clustering involves so many mergers that it seems almost impossible to
avoid forming a substantial classical bulge \citep{pee_nus_10}.

\subsection{The bulge structure may be younger than its stars }
A common misconception about the secularly evolved bar model is that it seems inconsistent with the old age of bulge stars that formed on a short timescale ($\sim$ 1 Gyr), as demonstrated by their $\alpha$-element enhancement. It is important to make a distinction between the assembly time of the bulge/bar and the age of the bulge stars; the Galactic bulge structure may be younger than its stars. The stars in our Galactic bar are older than most disk stars, but those stars could have formed over a short period of time long before the bar structure formed \citep{wyse_99,freema_08}. Their old age \citep[e.g.,][]{zoccal_etal_03, fulbri_etal_07,clarks_etal_08} is therefore not an argument against the internal secular evolution model.

\subsection{Could the vertical metallicity gradient be produced in a simple bar model?}
\label{sec:verticalgrad}

Although the general bulge morphology and kinematics are well explained by this simple and self-consistent model, a potential difficulty of the simple model is how to explain the observed vertical metallicity gradient \citep[e.g.,][]{zoccal_etal_08,gonzal_etal_11a}. Intuitively one may expect that any pre-existing vertical metallicity gradient should be erased due to mixing in the violent buckling process. Thus the vertical metallicity gradient has been suggested as the strongest evidence for the existence of a classical bulge in the Milky Way.

S10 proposed a plausible solution that an abundance gradient can still be produced within the context of secular bar/bulge formation if some of the vertical thickening is produced by resonant heating of stars that scatter off the bar \citep{pfe_nor_90}. If the most metal-poor stars are also the oldest stars, then they have been scattered for the longest time and now reach the greatest heights away from the disk plane.

Recently \cite{mar_ger_13} challenged the widespread belief that secularly evolved bar/bulge models cannot have metallicity gradients similar to those observed in our Galaxy. Using a similar boxy bulge/bar simulation as in S10, they were able to successfully reproduce the observed vertical metallicity gradient and longitude-latitude metallicity map similar to that constructed by \citet{gonzal_etal_13}, provided that the initial unbarred disk had a relatively steep radial metallicity gradient. One important assumption is that the pre-existing radial metallicity gradient is set up during the buildup of the precursor disk prior to bar formation. They proposed that if the Galactic bar/bulge formed rapidly from the precursor disk at early times, the violent relaxation may be incomplete during the bar and buckling instabilities, thus transforming radial pre-existing metallicity gradients to vertical gradients in the final boxy bulge.
They further proposed that the range of bulge star metallicities at various latitudes may be used to constrain the radial gradient in the precursor disk. In their study, they tagged particles with some metallicity and no chemical evolution was considered. If their result is confirmed in more detailed studies, then the vertical metallicity gradient is no longer a strong argument against the secularly-evolved bar/bulge model.

\subsection{Merits of the simple self-consistent bar/bulge model}

The main advantage of a simple self-consistent bar/bulge model such as the one in S10 lies in its simplicity; S10 tried to construct the simplest self-consistent $N$-body models that fit the BRAVA data, avoiding contrived models with too many free parameters. In addition, the S10 model is not just a static model but rather one that
evolved naturally to this state from simple initial conditions. The bar is self-consistently developed from a massive cold precursor disk embedded in a cored halo. So it has relatively few parameter to tweak, unlike the more complicated chemo-dynamical models.

Secondly, physical processes dictating the formation and evolution of the bar/bulge can be well understood. The formation of evolution of the bar is a natural outcome of two well-studied dynamical instabilities, namely the bar instability creating the bar and the subsequent firehose/buckling instability that thickens the bar vertically (see \citealt{sellwood_14} for a comprehensive review on these instabilities). The buckling instability also helps the Milky Way to develop an intriguing X-shaped structure (Section~\ref{sec:xshape}). The different components of the early inner Galaxy (thick disk, old and younger thin disks) probably become trapped dynamically within the bulge structure \citep{ness_etal_13b}.

Thirdly and probably most importantly, despite its simplicity, the best-fitting model of S10 also ties together several isolated results (e. g., photometric bar angle, kinematic bar angle, successful kinematic fits in the context of observed cylindrical rotation, reasonable bar length and bar pattern speed, resulting constraint on a classical bulge, the vertical metallicity gradient, and explaining the X-shaped structure) into a coherent picture, cemented by an $N$-body model in which the bar evolves naturally and without complicated fine-tuning. At current stage, it is still unrealistic to expect a full chemo-dynamical model to match all observed properties of the Galactic bulge. The simple bar model was designed to serve as a physically-motivated starting point, then one can gradually incorporate more complexities of the Milky Way bulge on top of it.

\begin{figure*}[!ht]
\centerline{\includegraphics[angle=0.,width=0.5\hsize]{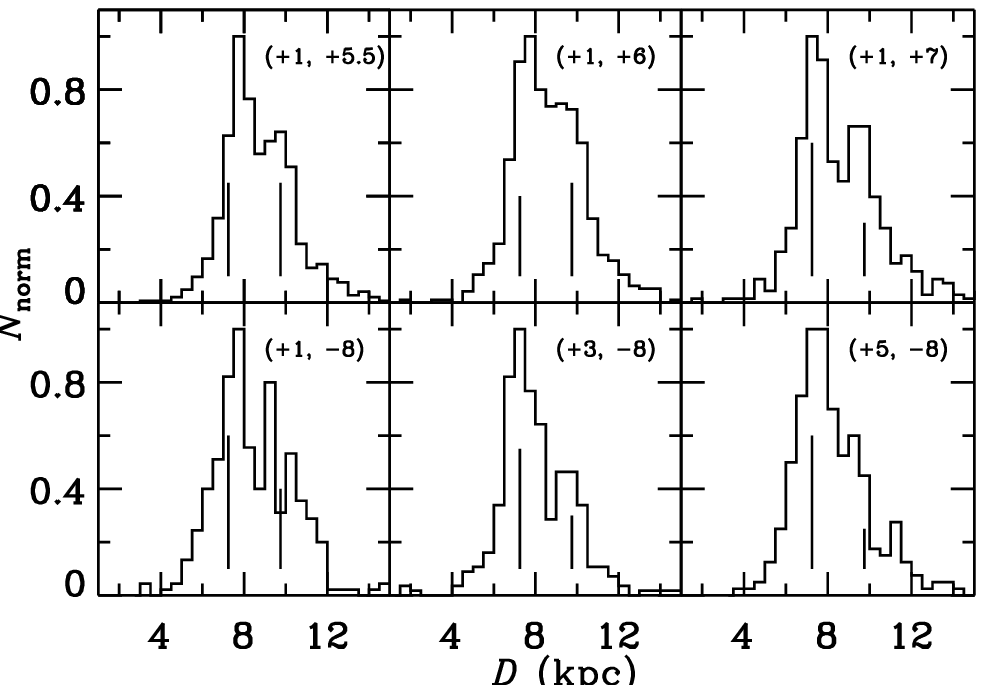}
\includegraphics[angle=0.,width=0.5\hsize]{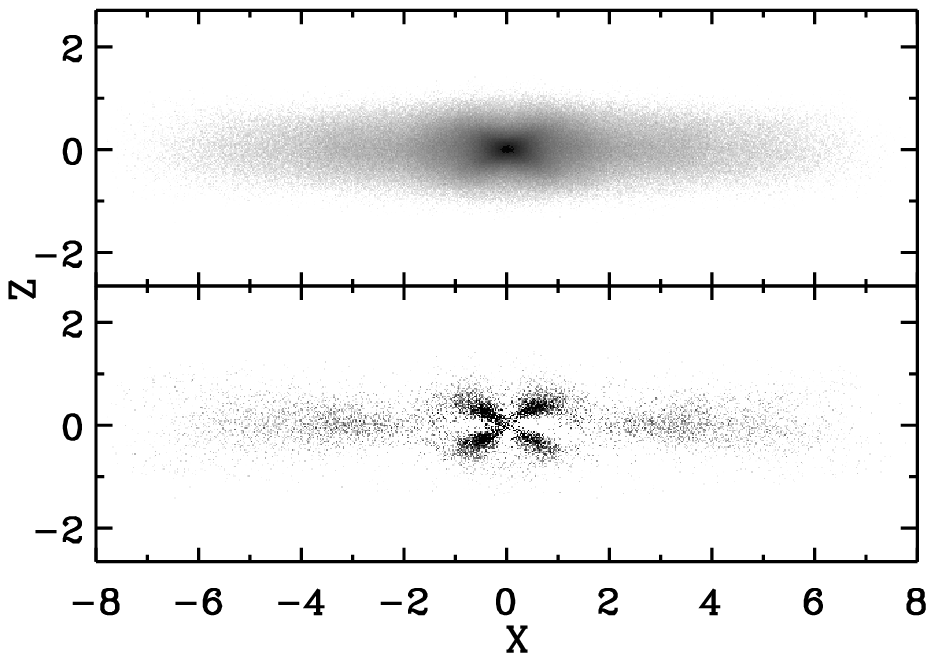}}
\caption{a). Left panels: Distance distributions of particles in the S10 model in the
Galactic bulge fields at a given longitude (top row) and latitude
(bottom row); b). Right panels: Demonstration of the X-shape structure with the
upper panel showing the side-on view of the bar in the S10 model and the
lower panel showing the residual after fitting and subtracting the underlying smooth
light contribution. The vertical X-shaped structure is highlighted in
this residual image. The length unit is $R_{\rm d} = 1.9 \rm \; kpc$. \textit{Reproduced from \citet{li_she_12} with permission.}}
\label{fig:xshape_lb}
\end{figure*}

\section{The X-shaped structure in the Galactic bulge}
\label{sec:xshape}

RCs are a good standard candle commonly used in the studies of the Galactic bulge. The apparent magnitude of RCs in high-latitude bulge fields shows a clearly bimodal distance distribution, often termed ``split red clump''. This bimodal distribution was discovered in the 2MASS data (MZ10, \citealt{saito_etal_11}) and the OGLE data \citep{nataf_etal_10}, then confirmed also in the ARGOS sample \citep{ness_etal_12}. The RCs seem to be distributed in a vertically-extended X-shaped structure. This X-shape was initially puzzling; it seemed difficult to explain them using a tilted naive ellipsoidal bar.

This X-shaped structure does not have a straight-forward explanation in
classical bulge formation scenarios, but it is a natural consequence of the bar buckling processes (see Section\ref{sec:bar}) if it is properly modeled \citep{combes_etal_90,raha_etal_91,bur_fre_99,athana_05,martin_etal_06}. \cite{li_she_12} analysed the same best-fitting Milky Way bar/bulge model in S10, and demonstrated that it can qualitatively reproduce many observations of the X-shaped structure, such as double peaks in distance histograms (MZ10, \citealt{nataf_etal_10}) and number density maps \citep{saito_etal_11}.

\citet{li_she_12} found that an X-shaped structure is clearly discernible in the inner region of the side-on view of the S10 best-fitting bar/bulge model (top panel of Fig.~\ref{fig:xshape_lb}b). They fitted and subtracted the underlying smooth component from the side-on bar model, then the X-shaped structure is highlighted in the bottom panel of Fig.~\ref{fig:xshape_lb}b \citep{li_she_12}. Fig.~\ref{fig:xshape_lb}a shows the distance distributions of particles towards the Galactic bulge, where double peak features similar to observations can be clearly identified. The bottom row shows that at a given latitude $b$ the peak positions are roughly constant. As the longitude decreases, the peak at larger distances becomes stronger. The top row shows that at a given longitude $l$ the separation of the two peaks increases as the line of sight is further away from the Galactic plane. These results nicely agree with the observations of the X-shaped structure (MZ10). Along the bar major axis, the end-to-end separation between the inner two edges of the X-shaped structure is $\sim$2 kpc. For the outer two edges, the end-to-end separation is $\sim$4 kpc. The size of the X-shaped structure along the bar is estimated by averaging the two separations, which yields $\sim$3 kpc. It is worth noting that the value is much less than the bar's full length ($\sim$8 kpc). Similarly, the end-to-end vertical separation between the inner two edges of the X-shaped structure is $\sim$1.2 kpc. For the outer two edges in the vertical direction, this separation is $\sim$2.4 kpc. Therefore the vertical size of the X-shaped structure in the S10 model is $\sim$1.8 kpc. By
summing up the pixels with positive values in the X-shaped region, \citet{li_she_12} estimated that the light fraction of this X-shaped structure relative to the whole boxy bulge region is about 7\%. It is still uncertain how much mass is contained in the X-shape. More sophisticated analysis by \cite{portai_etal_15a}, based on the reconstructed volume density of the Galactic bulge from Vista Variables in the Via Lactea survey (VVV) \citep{weg_ger_13}, suggests an off-centered X-shape enclosing about 20\% of the bulge mass. Recently \cite{nataf_etal_15} studied the X-shape properties based on OGLE-III observations, and also found good agreement with the models in S10 and \citet{ness_etal_12}.

\citet{li_she_12} also demonstrated that the X-shaped structure becomes
nearly symmetric with respect to the disk plane about 2 Gyr after the buckling instability gradually
saturates.  The observed symmetry (MZ10) probably implies that the X-shaped structure in the Galactic
bulge has been in existence for at least a few billion years.

Based on the ARGOS sample, \cite{ness_etal_12} found that the X-shaped structure is mainly composed of the metal-rich stars rather than the metal-poor ones (also see \citealt{uttent_etal_12,gilmor_etal_12}). However, this conclusion is questioned by \cite{nataf_etal_14} who demonstrated that there may be a bias of metallicity on the RCs distance determination. In the future larger and unbiased samples are required to answer this question unambiguously.

The existence of the X-shaped structure in our Milky Way provides extra evidence that the Galactic bulge is shaped mainly by internal disk dynamical instabilities instead of mergers, because no other known physical processes can naturally develop such a structure. \citet{deprop_etal_11} studied the radial velocity and abundances of bright and faint RCs at $(l, b) = (0^\degree,-8^\degree)$, and found no significant dynamical or chemical differences. This may suggest that the two RCs indeed belong to the same coherent dynamical structure, which can be naturally made in the formation of the bar/boxy bulge.

Orbital structure studies are essential for understanding the properties of the X-shaped structure, which is the outcome of the collective buckling instability. The backbone orbits of a three-dimensional buckled bar are the $x_1$ tree, i.e., the $x_1$ family plus a tree of three-dimensional families bifurcating from it \citep{pfe_fri_91}. It is widely believed that the X-shaped structure may be supported by orbits trapped around the three-dimensional $x_1$ family (also known as banana orbits due to their banana shape when viewed side-on) \citep[e.g.,][]{patsis_etal_02,skokos_etal_02a,pat_kat_14a}. But this picture was recently challenged by \citet{portai_etal_15b} who argued that the X-shaped bulge is mainly supported by brezel-like orbits instead. \citet{qin_etal_15} also found that stars in the X-shaped bulge do not necessarily stream along simple banana orbits. Clearly, more studies on the orbital structure and vertical resonant heating \citep[e.g.][]{quille_etal_14} are desired to make more specific predictions for the Milky Way.

The detailed kinematics of the near (bright) and far (faint) sides of the X-shape may be a useful tool to probe the underlying orbital structure. Several observational studies have explored the X-shape kinematics.
With about 300 RCs in $(0^\degree, -6^\degree)$, \cite{vasque_etal_13} found a weak anti-correlation between the longitudinal proper motion and radial velocity in both bright and faint RCs. However, they found no significant correlation between the latitudinal proper motion and the radial velocity. These results were interpreted as possible streaming motions along the X-shaped arms. In bulge fields at $b \sim 5^\degree$, \cite{polesk_etal_13} reported the asymmetric mean proper motion difference between the near and far sides in both $l$ and $b$ directions; this difference is linear for $-0.1^\degree < l < 0.5^\degree$, but roughly constant for $-0.8^\degree < l < -0.1^\degree$. The linear part was attributed to the streaming motions in the X-shape.

Numerical simulations can provide comprehensive understanding of the X-shape kinematics. The kinematic imprints of the X-shape onto the Galactic bulge observations were interpreted in \cite{gardne_etal_14} based on their $N$-body models; they found a coherent minimum along $l = 0^\degree$ in the mean radial velocity difference between the near and far sides, which is absent in other velocity components and all the velocity dispersions. \citet{qin_etal_15} systematically explored the kinematics (both the radial velocity and the proper motion) of the X-shape in the S10 model. Along $l = 0^\degree$, the near and far sides of the bar/bulge show excess of approaching and receding particles, reflecting the coherent orbital motion inside the bar structure. \citet{qin_etal_15} found only very weak anisotropy in the stellar velocity within the X-shape, hinting that the underlying orbital family of the X-shape may not be dominated by simple banana orbits. Contrary to \cite{polesk_etal_13}, \citet{qin_etal_15} found that the proper motion difference between the near/far sides of the X-shape may not be used to constrain the bar pattern speed. They also confirmed that the Galactic center may be located by fitting the arms of the X-shape, which was originally proposed by \cite{gardne_etal_14}.

\begin{figure*}[!ht]
\centerline{ \includegraphics[angle=0.,width=\hsize]{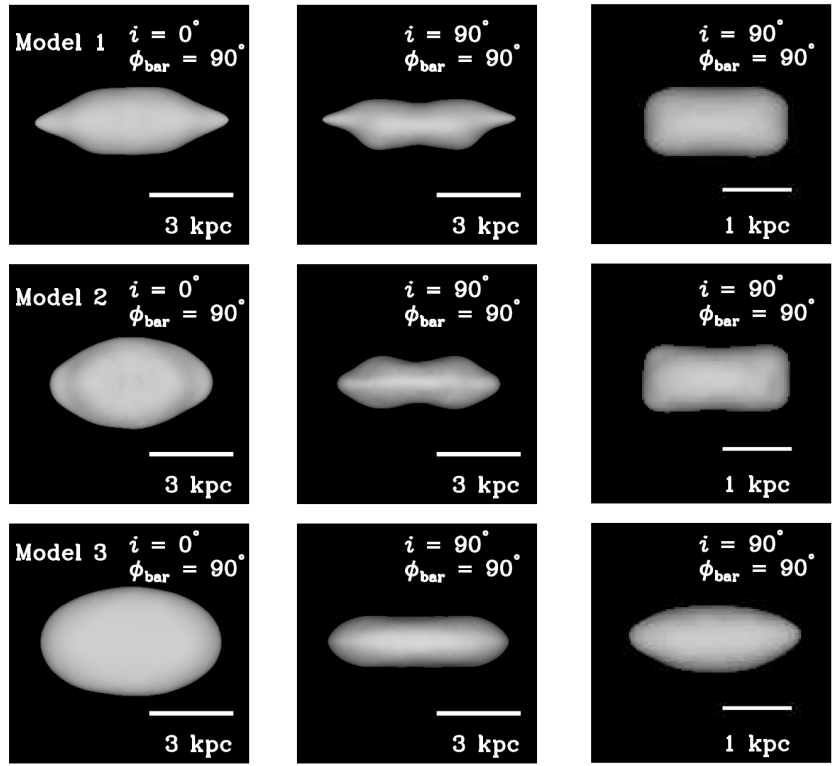}}
\caption{Isodensity surfaces of three $N$-body bars with different buckling strengths, namely, Model 1 (top row, strongest buckling), Model 2 (middle row, intermediate buckling, the S10 model) and Model 3 (bottom row, negligible buckling). The left and middle columns show the face-on and side-on appearance of the bar at large scale ($\sim$ 3 kpc). The right column shows the side-on shape of the isodensity surfaces at small scale ($\sim$ 1 kpc).}
\label{fig:3dx}
\end{figure*}

It is important to keep in mind that the true 3D shape of the X-shape should not be visualized as simple as a letter ``X" with four or eight conspicuous arms sticking out. Fig.~\ref{fig:xshape_lb} may give you such an impression because human eyes are easily biased to pick out the small-scale density enhancement. The true 3D shape or structure of iso-density surfaces should really be more like a peanut. This is demonstrated in \citet{li_she_15} who estimated the 3-D volume density for three $N$-body simulations with the adaptive kernel smoothing technique \citep{silver_86, she_sel_04}; these three bars have undergone different amplitudes/strengths of buckling instability. Fig.~\ref{fig:3dx} shows clearly that a buckled bar is composed of three components with increasing sizes: a central boxy core, a peanut bulge and an extended thin bar (Fig.~\ref{fig:3dx}).  The true 3D structure of the X-shape is actually more peanut-shaped (Fig.~\ref{fig:3dx}), but the peanut-shaped bulge can still qualitatively reproduce the observed bimodal distance distributions that were used to infer for the X-shape. Our visual perception of seeing an ``X" is enhanced by the pinched concave shape of the inner peanut structure.

\section{More sophisticated chemo-dynamical models of the Galactic Bulge}
\label{sec:chemodyn}

Despite its simplicity, the self-consistent Milky Way bar/bulge model described in Section~\ref{sec:bar} is successful in explaining many aspects of the Galactic bulge, such as the boxy shape, the stellar kinematics, the X-shape, and gives reasonable bar angle, length, and the pattern speed. It may also explain, in principle, the vertical metallicity gradient (\citealt{mar_ger_13} and Section~\ref{sec:verticalgrad}).  However, in order to match the full sophisticated distribution of the stellar populations and chemical composition in the Galactic bulge, more realistic chemo-dynamical simulations must  eventually supersede the simple $N$-body disk simulations (Section~\ref{sec:bar}) in the future. It is not trivial to model gas properly in simulations, because we need to treat many complicated micro-physicical processes of the multi-phase medium with simplified prescriptions. In particular, there are still considerable uncertainties in how to model the star formation and feedback processes.

Most chemo-dynamical models are motivated to explain the vertical metallicity abundance of the Galactic bulge, which was thought to be the main challenge for the secularly evolved bar/bulge model. However, as we discuss in Section~\ref{sec:verticalgrad}, the vertical metallicity gradient may be explained by the simple bar/bulge model as well. With additional mass components in the more complicated chemo-dynamical simulations, and many more free parameters, the vertical metallicity gradient may also be reproduced in these simulations.

\subsection{Two-component disk scenario}

\citet{bek_tsu_11b} reported that their pure thin disk model failed to reproduce the vertical metallicity gradient in the Galactic bulge, which is opposite to the conclusion reached in \citet{mar_ger_13}.  This difference is due mainly to the different initial setup in the two studies.  Unlike \citet{bek_tsu_11b}, \citet{mar_ger_13} did not try to link the initial radial metallicity profile to the metallicities of the present-day Galactic disk near the Sun, since the buckling instability in the Milky Way must have occurred long ago when the outer disk was only incompletely assembled.

Nevertheless, in order to explain the observed metallicity gradient \cite{bek_tsu_11b} proposed a two-component disk scenario where the bulge is formed from a disk composed of thin and thick disks. In this scenario, the first thin disk was disturbed and heated into a thick disk via a minor merger with a dwarf galaxy at early times. A subsequent thin disk is gradually built up with the inner part developing a bar structure. Since more metal-poor stars at higher latitude originate from the already dynamically hotter thick disk, which was not strongly influenced by vertical mixing of the later bar, they  are able to stay in-situ for much longer and keep the metallicity low at high vertical distance. Consequently, a vertical metallicity gradient
of the bulge can be produced. A similar model was also suggested in \cite{dimatt_etal_15}. This scenario may help explain the similarity in stellar populations between the bulge and the thick disk \citep{melend_etal_08,alvesb_etal_10,bensby_etal_11}. However, it is still not entirely clear whether or not the bulge stars are a distinct population with different kinematics and compositions from the thick disk stars \citep[e.g.,][]{lecure_etal_07,fulbri_etal_07,min_zoc_08}.

\subsection{Clump-origin bulges}

Clump-origin bulges form through mergers of clumps in a primordial galactic disk. There have been attempts to link the Milky Way bulge with a clump-origin bulge. More discussions of clump-origin bulges may be found in Chapter~6.2 of this book.

Primordial galaxies are expected to be highly gas-rich in early stage of the disk formation, and massive star-forming clumps may form by gravitational instabilities. The clump can spiral into the center of the galaxy rapidly by dynamical friction, then merge to form a central bulge component \citep{noguch_99, immeli_etal_04a, immeli_etal_04b,bourna_etal_07,elmegr_etal_08,ino_sai_12}. These results are consistent with the clumps in the chained galaxies observed at high redshift \citep{elmegr_etal_04}.

The reported properties of the clump-origin bulges still differ considerably in various numerical simulations. \cite{elmegr_etal_08} found the similarity between clump coalescence and major galaxies mergers in terms of orbital mixing. In their simulations, these giant star forming clumps migrated towards the galaxy center within a few dynamical time scales to form a classical bulge, that is thick, rotates slowly and follows a S\'{e}rsic density profile with large index (i.e. close to the $R^{1/4}$ law profile). \cite{ino_sai_12} performed high resolution $N$-body/SPH simulations of clump-origin bulges by assuming a collapsing gas sphere embedded in an NFW halo. They found that their bulge has many properties indicating for a pseudobulge, such as a nearly exponential surface density profile, a barred boxy shape and strong rotation. They also found that this bulge consists of old and metal-rich stars, similar to stars in the Galactic bulge. However, the clump-origin bulges made in these studies all failed to produce the defining characteristic of the Milky Way bulge, i.e., the strong cylindrical rotation pattern \citep{howard_etal_09,shen_etal_10}. The relevance of clump-origin bulges to the Milky Way needs to be examined in greater depth in the future.

\subsection{The Milky Way bulge formation in the cosmological setting}

Ideally one would prefer to simulate the full evolution history of Milky Way formation with cosmological initial conditions. Such simulations will simulate the formation of the Galactic bulge in addition to thin and thick disks. The ambitious full chemo-dynamical simulations combine 3-D hydrodynamical simulations with the calculation of chemical enrichment, and obtain the positions, ages, chemical compositions, kinematics of all star particles. Such detailed predictions can be compared to observations in large surveys. However, cosmological chemo-dynamical simulations are very computationally expensive, and much realism was still fudged in sub-grid physics.

With improved star formation and feedback models, cosmological disk simulations have made progress in making late-type disk galaxies in recent years. These simulations often use strong feedback models to prevent overproduction of stars at early times \citep[see the recent review by][]{gerhar_14}. \citet{obreja_etal_13} analyzed and compared the bulges of a sample of $L_*$ spiral galaxies in hydrodynamical simulations in a cosmological context. They found that the bulges show an early starburst-collapse fast phase of mass assembly, followed by a second phase with lower star formation rate, driven by disk instabilities and/or minor mergers. They suggested that one may associate the old population formed during the first rapid phase with a classical bulge, and the young one formed during the slow phase with a pseudobulge. The young population is more oblate, generally smaller, more rotationally supported, with higher metallicity and less alpha-enhanced than the old one. \citet{guedes_etal_13} followed the formation and evolution of the pseudobulge in the Eris hydrodynamic cosmological simulation with a very high resolution. Their pseudobulge was built from the inner disk, and composed of mainly old stars that formed in the first step in the inside-out formation of the baryonic disk. So far these cosmological simulations \citep{guedes_etal_13,kobaya_14} still do not have a boxy/peanut-shaped bulge as in the Milky Way.  Cosmological simulations need a huge dynamic range to model the bulge and halo simultaneously, since the bulge is only $\sim 1\;\kpc$ in size, and the halo is around $\sim 200\;\kpc$. So the spatial resolution required to resolve structures in the bulge may be too high. There is still a long way to go for the current chemo-dynamical simulations to explain the full evolution history of the Milky Way's bulge.

The ultimate goal of chemo-dynamical simulations is not only to reproduce the final results similar to observations, but also to make clear which process is responsible for each particular aspect of observations, which may require a large set of parameter studies. Only until we identify the basic factors contributing to the more complex features can we claim to understand all the physics involved in the formation and evolution of the Milky Way.

\section{Other modelling techniques}
\label{sec:othertech}

\subsection{Bar properties constrained from gas kinematics}

The presence of a bar structure in the Mliky Way was first hinted by the non-circular gas kinematics \citep[e.g.,][]{devauc_64}. One may further use the features in the asymmetric gas flow pattern to infer the properties of the Galactic bar/bulge. Non-circular motions of atomic and molecular gas in the Galaxy are commonly presented in the $l-v$ diagram, which shows the distribution of gas emission line intensity as a function of Galactic longitude and line-of-sight velocity since distances to individual gas clouds are difficult to measure \citep[e.g.,][]{bur_lis_93, dame_etal_01}.  The high density features in the $l-v$ diagram represent the over-dense regions of gas distribution driven mainly by the large-scale non-axisymmetric structures such as the Galactic bar and spiral arms. The features in the $l-v$ diagram, due to their unknown distances, must be interpreted through gas dynamical models, and they can provide important constraints on the properties of the bar and spiral arms.

There have been many hydrodynamic models of the gas flow in barred potentials derived from COBE near-infrared maps or star counts \citep[e.g.,][]{eng_ger_99, bissan_etal_03,rod_com_08}, or in self-consistent barred models \citep[e.g.,][]{fux_99, baba_etal_10}. These models have been able to reproduce many of the strong features in the $l-v$ diagram, even though no model was able to provide a good match to all the observed features. \cite{fux_99} modelled gas dynamics with 3D $N$-body + SPH simulations. He found that the gas flow driven by a self-consistent bar is asymmetric and non-stationary. Snapshots at some specific times can qualitatively reproduce the observed $l-v$ diagrams of HI and CO. His model could reproduce the connecting arms, which represent the dust lanes as a result of strong bar-driven shocks, 3-kpc arm and 135-$\kms$ arm (both arms are emanating from the ends of the bar). Based on his modeling results, he found that a  bar angle of $25^\degree \pm 4^\degree$,  a bar co-rotation radius of 4.0-4.5 $\kpc$, and a bar pattern speed of $\sim 50\;\kmskpc$, consistent with his stellar modelling results \citep{fux_97}.

\cite{eng_ger_99} studied gas dynamics in a rotating gravitational potential of the deprojected COBE near-infrared bar and disk. In their models, gas formed a pair of shocks at the leading side of the bar and a nuclear ring, typical of the gas flow pattern in a barred potential. Many observed gas dynamical features could be found in their models, such as the four-armed spiral structure, the 3-kpc arm, the terminal velocity curve (i.e., the envelope line of the $l-v$ diagram), tangent points, although some features are not exactly identical to those in observations. They predicted a bar pattern speed of $\sim 60\;\kmskpc$ and a bar angle of about $20-25^\degree$, similar to the results from \citet{fux_99}.
\cite{bissan_etal_03} simulated the gas flow with an updated potential further constrained by COBE maps and clump giant star counts in several bulge fields. They adopted separate pattern speeds for the bar and spiral arms, and suggested that such a configuration would help to make spiral arms pass through the bar co-rotation radius where the spiral arms dissolve in the single pattern speed simulations.
The 3-kpc arm and its far-side counterpart in this work cannot be reproduced very well unless they use a massive spiral arm potential. They predicted a bar pattern speed of $\sim 60\;\kmskpc$ and spiral arm pattern speed of $\sim 20\;\kmskpc$, similar to  previous results.

The bar pattern speed derived from gas dynamics may have some degeneracy with the bar size; a large bar length coupled with a lower bar pattern speed may work quite well to match gas observations. \citet{liz_etal_15} modelled gas flow pattern for the Milky Way using grid-based hydrodynamical simulations. Their basic bar potential was from an $N$-body model constrained by the density of bulge red clump stars \citep{portai_etal_15a,weg_ger_13}. They found that a low pattern speed model for the Galactic bar may work, and the best model can give a better fit to the $l-v$ diagram than previous high pattern speed simulations.

\subsection{Orbit-based and particle-based modelling}

Although numerical simulations such as $N$-body can offer the full evolutionary history from plausible initial conditions, they also have weaknesses.  Numerical simulations are inflexible in the sense that a lot of trials are required to reproduce the desired results by adjusting the initial conditions --- this becomes very challenging especially for more sophisticated chemo-dynamical simulations which often rely on fudged sub-grid physics and are computationally costly, thus limiting the systematic exploration of parameter space to match the observational results.

Other modelling approaches such as the \textbf{Schwarzschild} and \textbf{made-to-measure} methods are complementary to $N$-body simulations. Both methods are good at steering models to match the desired final results.

The basic principle of the Schwarzschild orbit-superposition method is given in \citet{schwar_79}. One first computes and finds the typical orbital families in a potential arising from an assumed triaxial density distribution. The time-averaged density along each orbit in a lattice of cells spanning the volume of the model is simply the total cumulative time spent by that orbit in each cell. Then linear/quadratic  programming techniques are used to find the non-negative weights of each orbit to fit the assumed mass distribution and the kinematical data to achieve self-consistency. There is no unique solution for the orbital weights, but one may seek to maximize some objective function for a balance of $\chi^2$ minimizations and smoother phase-space distributions \citep[e.g.,][]{ric_tre_88,thomas_etal_05}. The stability of the constructed model is also not guaranteed, and generally needs to be tested with an $N$-body code. The weighted orbital structure resulting from the Schwarzschild modeling can offer important clues to the formation history of system.

Based on the Schwarzschild orbit-superposition method, \citet{zhao_96} developed the first 3D rotating bar model that
fitted the density profile of the COBE light distribution and kinematic data at Baade's window. His model was constructed with 485 orbit building blocks, and little stellar kinematic data were available to explore the uniqueness of this steady-state model. \citet{hafner_etal_00} extended the classical Schwarzschild technique by combining a distribution function that depends only on classical integrals with orbits that respect non-classical integrals, i.e., Schwarzschild's orbits were used only to represent the difference between the true galaxy distribution function and an approximating classical distribution function. They used the new method to construct a dynamical model of the inner Galaxy with an orbit library that contains about 22,000 regular orbits. For definiteness, they assumed a bar angle of $20^\degree$ and bar pattern speed of $60\;\kmskpc$. The model reproduced the 3D mass density obtained through deprojection of the COBE surface photometry, and the then-available kinematics within the bar corotation radius (3.6~kpc).

\citet{wang_etal_12} extended the Schwarzschild implementation of \citet{zhao_96} and applied it with the extra kinematic constraints from the full BRAVA dataset \citep{kunder_etal_12}. Using $\chi^2$ minimization, their best-fitting Galactic bar model has a pattern speed of $60\;\kmskpc$, a disk mass of $10^{11}\Ms$ and a bar angle of $20^\degree$ out of 36 models varying these parameters.
Compared to \citet{zhao_96}, \citet{wang_etal_12} can better reproduce the average radial velocity and the surface brightness distribution. However, their model over-predicted the longitudinal proper motions compared to the observed values. $N$-body tests showed that the model was stable only for a short period of 0.5 Gyr. They suspected the instability arises because no self-consistency was imposed for the disk outside 3 kpc.
\citet{wang_etal_13} made further tests of their implementation using the $N$-body bar model in \citet{shen_etal_10}, with the hope to recover the given bar pattern speed and the bar angle in \citet{shen_etal_10}. They concluded that BRAVA radial velocities alone do not constrain well the bar angle and/or the pattern speed using their Schwarzschild implementation, and the observed proper motions may help to reduce the model degeneracy. Their method appeared to over-fit the BRAVA data points, indicating the implemented smoothing of the phase space distributions may need improvement.

Building a complete and representative orbit library in bars, essential for the Schwarzschild method, is non-trivial. It is necessary to explore systematically which of the many possible orbit families could contribute to a self-consistent model, and to understand how they are affected by properties of the bar potential. For example, Schwarzschild modelling may help elucidate the actual orbital compositions supporting the prominent X-shaped/peanut-shaped structure in the Galactic bulge. There seem to be a large number of irregular/chaotic orbits in bars \citep[e.g.,][]{wang_etal_12,pat_kat_14a,pat_kat_14b}. We need to have a better idea on the fraction of chaotic orbits, classify them into major families, and identify the location of resonant orbits.

Uniqueness of the solution is also an undesired feature in the Schwarzschild modeling of the Galactic bar. There is considerable freedom to reproduce the existing data by different sets of orbital weights. More systematic studies are needed to explore how many different combinations of orbits still reach the same goodness-of-fit for a given potential, and how much the properties of the final bar can vary but still satisfy the same observational constraints.


Unlike the Schwarzschild method, \textbf{made-to-measure (M2M)} method \citep{sye_tre_96} slowly adjusts the weights of the particles in an $N$-body system, instead of the orbital weights as in the Schwarzschild method, as particles proceed in their orbits until the time-averaged density field and other observables converge to a prescribed observational value.  The particle weights are adjusted through a weight evolution equation according to the mismatch between the model and target observables. In the Schwarzschild method orbits are first separately integrated in a fixed potential and then superimposed, whereas in the M2M method the two steps are merged --- orbits are integrated and the weights of particles are adjusted at the same time, thus eliminating the need for an orbit library. Also M2M can allow us to dynamically adjust the potential while Schwarzschild does not.

The M2M method was first applied to construct a dynamical model for the barred bulge and disk model of the Milky Way in \citet{bissan_etal_04}. Since then the M2M method has been continuously improved in various implementations, and has gained growing interests in the dynamical modelling of galaxies, especially spheroidals and  early-type galaxies \citep[e.g.,][]{delore_etal_07,delore_etal_08,delore_etal_09,dehnen_09,lon_mao_10,lon_mao_12,mor_ger_12,hun_kaw_13,hunt_etal_13}.
A modified $\chi^2$M2M was implemented to improve the algorithm to model both the density and kinematic data,  account for observational errors and seeing effects, and incorporate a maximum-likelihood technique to account for discrete velocities \citep{delore_etal_07,delore_etal_08}. \cite{lon_mao_12} applied their design and implementation of the M2M method \citep{lon_mao_10} to 24 SAURON galaxies previously analysed by \citet{cappel_etal_06}, and found generally good agreement between M2M and Schwarzschild methods in determining the dynamical mass-to-light ratio.

\cite{long_etal_13} constructed a M2M model of the Galactic bar/bulge constrained by the BRAVA kinematics \citep{kunder_etal_12}. They took the $N$-body model in \cite{shen_etal_10} as initial condition and the target luminosity density. They ran a suite of 56 models with different pattern speeds and bar angles in search of the best-fitting one. Their best-fitting model recovered the bar angle and pattern speed of the
\citet{shen_etal_10} $N$-body model, and reproduced both the mean radial velocity and radial velocity dispersion of the BRAVA data very well. Since they used BRAVA results as kinematic constraints and the \cite{shen_etal_10} model as the target luminosity density, this work is actually a cross-check between the direct $N$-body modelling and the M2M method.
\cite{hun_kaw_13, hunt_etal_13} modelled disk systems with their own design and implementation of the M2M method (particle-by-particle M2M method, or PRIMAL). They showed that PRIMAL can recover the radial profiles of the surface density, velocity dispersions, the rotational velocity of the target disks, the apparent bar structure and the bar pattern speed of the bar.

Recently, \cite{portai_etal_15a} constructed dynamical models of the Galactic box/peanut bulge, using the 3D density of red clump giants  \citep{weg_ger_13} and BRAVA kinematics as observational constraints. They tried to match the data using their M2M code \citep{delore_etal_07,delore_etal_08}, starting with $N$-body models for barred disks in different dark matter haloes. In this work, they estimated the total dynamical mass (including both stellar and dark matter halo) as $1.84\pm0.07\times 10^{10} \Ms$ inside the rectangular box of  $\pm2.2 \times \pm1.4 \times \pm1.2 \; \kpc$).  They used BRAVA kinematical data to constrain the models, but the  proper motion data \citep{ratten_etal_07a} are used only as a check of their modelling. Their models tend to be more anisotropic than the data.
Given the different significance of the disk component in their initial conditions, their five models sample a wide range of pattern speeds of the final bar structures, ranging from $R\sim1.08$ to $1.80$, where $R=R_{\rm CR}/R_{bar}$ is the dimensionless parameter characterizing the pattern speed of the bar. They got a scaled value of $25$ to $30\;\kmskpc$, significantly lower than previous estimates with various methods ($40$ to $60\;\kmskpc$). If confirmed, this puts the Galactic bar among slow rotators  ($R \ge 1.5$).

In summary, both the Schwarzschild and M2M methods are designed to steer models with pre-determined initial conditions towards prescribed observed results. M2M models include aspects of both Schwarzschild methods and regular $N$-body simulations. When the gravitational potential of the target system is held fixed, the searching process for a distribution of particle weights is closely related to that for a distribution of orbital weights in Schwarzschild¡¯s method to fit the same observational constraints. Conversely, when the adjustment of particle weights is switched off and the potential is allowed to evolve, M2M particle codes can reduce to $N$-body simulations \citep{gerhar_10_HiA}. On the other hand, both types of modelling only construct self-consistent equilibrium models, but they do not tell us how the Galaxy evolved into the current equilibrium configuration from what initial conditions. The most important value of these modelling techniques is that they give us predictive power and suggest further observational tests to confirm or constrain the structural parameters of the bar/bulge.

\section{Summary and future outlook}
\label{sec:summary}

Understanding the structure and formation of our Milky Way bulge is nontrivial, mostly because of our location in the disk plane and the severe dust extinction in the optical band. Near infrared images from the COBE satellite presented the first clear evidence of a boxy bulge in the Galaxy. Recent large dedicated surveys have allowed a large number of bulge stars to be studied individually in detail. Most bulge stars are very old with a wide range of metal abundances, and they formed earlier than most disk stars on a rapid formation time scale. Observationally, we still need to better understand the bulge metallicity components identified by \citet{ness_etal_13a} and their kinematic signatures \citep{ness_etal_13b}. Future large surveys will also need to settle whether or not the recently identified X-shaped structure is populated mainly by metal-rich stars, instead of metal-poor ones \citep{nataf_etal_14}.

The Galactic bulge contains crucial information about the formation of evolutionary history of the Milky Way.
To unravel these clues theoretical modelling of the bulge is essential. Here we have reviewed recent advances in modelling the Galactic bulge with $N$-body, chemo-dynamical simulations, and other modelling techniques. The main body of the Milky Way bulge appears to be a buckled/thickened bar seen somewhat end-on, as hinted from its asymmetric boxy shape. One can construct a fully evolutionary bar model that matches many properties of the Galactic bulge reasonably well (Section~\ref{sec:bar}). The dynamical evolution of the bar/bulge was driven mainly by two consecutive disk instabilities.  The bar forms naturally from a cold massive precursor disk via the well-known bar instability. Shortly after its formation, the bar suffers from a vigorous buckling instability, and becomes a thickened structure that appears boxy or peanut-shaped when seen edge-on.  Such a model self-consistently evolves from plausible simple initial conditions, and is successful in explaining many aspects of the Milky Way bulge, such as the excellent match to the kinematics of the whole bulge, the X-shaped structure that naturally arises in the bar buckling process, reasonable bar angle and other bar parameters consistent with independent structural analysis, and the metallicity map.

This simple model provides a promising starting point, but there are still many open questions to be answered in more sophisticated chemo-dynamical models. For example, what is the exact mass fraction of a possible classical bulge hidden underneath the dominant boxy bulge; how the fossil record of the early inner Galaxy (thick disk, old and younger thin disks) is mapped into the bulge structure, i.e., how to better understand the bulge metallicity components identified by \citet{ness_etal_13a, ness_etal_13b} in the ARGOS survey, and their correlation with the kinematics; how the strongly barred X-shaped structure is populated preferentially by metal-rich stars; is there a solid connection between the bulge and the thick disk \citep{bek_tsu_11b,dimatt_etal_15}?

Ongoing and upcoming large surveys will undoubtedly shed new light on the Milky Way bulge. \textit{Gaia} will provide accurate parallaxes and proper motions of about 20 million stars along all the lines of sight towards the bulge \citep{robin_etal_05}. Complementary to the \textit{Gaia} mission, ongoing ground based surveys such as APOGEE, VVV, \textit{Gaia}-ESO, GIBS will also provide us huge amount of high-resolution spectroscopic data. With the large influx of data and the improvements in theoretical models, we are poised to make greater progress in putting together all puzzle pieces of the Milky Way bulge.

\acknowledgement We thank Shude Mao, Dimitri Gadotti, Eija Laurikainen, Zhi Li, Richard Long, Martin Smith, and Yu-jing Qin for helpful discussions and comments. The research presented here is partially supported by the 973 Program of China under grant no. 2014CB845700, by the National Natural Science Foundation of China under grant nos.11333003, 11322326, 11403072, and by the Strategic Priority Research Program  ``The Emergence of Cosmological Structures'' (no. XDB09000000) of the Chinese Academy of Sciences. ZYL is grateful for the support from Shanghai Yangfan Talent Youth Program (No. 14YF1407700). This work made use of the facilities of the Center for High Performance Computing at Shanghai Astronomical Observatory.

\bibliography{shen}
\bibliographystyle{mn2e}

\end{document}